\documentclass{elsart3p}

\usepackage{natbib}

\usepackage{graphicx}

\usepackage{amssymb}

\newcommand{\Mwd}{\mbox{$M_\mathrm{wd}$}}

\newcommand{\Rwd}{\mbox{$R_\mathrm{wd}$}}

\newcommand{\Rout}{\mbox{$R_\mathrm{out}$}}
\newcommand{\Rin}{\mbox{$R_\mathrm{in}$}}

\newcommand{\Msun}{\mbox{$\mathrm{M_{\odot}}$}}
\newcommand{\Rsun}{\mbox{$\mathrm{R_{\odot}}$}}
\newcommand{\Teff}{\mbox{$T_\mathrm{eff}$}}

\newcommand{\Mdot}{\mbox{$\dot{M}$}}

\def\apgt{\ {\raise-.5ex\hbox{$\buildrel>\over\sim$}}\ }
\def\aplt{\ {\raise-.5ex\hbox{$\buildrel<\over\sim$}}\ }
\newcommand {\be} {\begin{equation}}
\newcommand {\ee} {\end{equation}}

\begin{document}

\begin{frontmatter}



\title{Spectra of accretion discs around white dwarfs}


\author[albl1,albl2]{Irit Idan,}
\author[albl3,albl4]{Jean-Pierre Lasota,}
\author[albl5]{Jean-Marie Hameury,}
\author[albl2]{Giora Shaviv}

\address[albl1]{Rafael, 31021 Haifa, Israel}
\address[albl2]{Department of Physics, Technion-Israel Institute of Technology, 32000 Haifa, Israel}
\address[albl3]{Institut d'Astrophysique de Paris, UMR 7095 CNRS, UPMC Univ Paris 06,
               98bis Bd Arago, 75014 Paris, France}
\address[albl4]{Astronomical Observatory, Jagiellonian University, ul. Orla 171, 30-244 Krak\'ow, Poland}
\address[albl5]{UMR 7550 du CNRS, Observatoire de Strasbourg, 11 rue de l'Universit\'e, F-67000 Strasbourg, France}

\begin{abstract}
We present spectra of accretion discs around white-dwarfs calculated
with an improved and updated version of the Shaviv \& Wehrse (1991)
model. The new version includes line opacities and convective energy
transport and can be used to calculate spectra of hot discs in
bright systems (nova--like variables or dwarf novae in outburst) as
well as spectra of cold accretion discs in quiescent dwarf novae.

\end{abstract}


\end{frontmatter}

\section{Introduction}
\label{sec:intro}

In weakly magnetized cataclysmic variable stars (CVs) matter lost by
a Roche-lobe filling low-mass companion forms an accretion disc
around the white-dwarf primary. In bright systems such as nova-like
binaries and dwarf-novae in outburst the disc is the dominant source
of luminosity but even in quiescence the disc emission might provide
crucial information about the physics of accretion, in particular
about the mechanisms transporting angular momentum and releasing
gravitational energy.

Until now one could calculate only spectra for hot accretion discs
\citep[see][and references therein]{wh98}. In such an approach the
disc is divided into rings whose vertical structure is
calculated using the program TLUSDISK \citet{hubeny90,hubeny91}. A
different way of solving radiative transfer equations in accretion
discs was presented by \citet[][hereafter SW]{sw91}. Also, in their
approach one could apply it to hot discs only. The preliminary attempt
to apply the SW code to cold quiescent discs in \citet{Idan1} was
not conclusive. The main obstacle in solving the vertical structure
of cold accretion discs is the presence of convection-dominated
zones. \citep[The description of the quiescent disc of SS Cyg by][neglects convection]{kromer}.

The absence of spectral models for cold quiescent discs is the
reason why fitting models to observations can be a frustrating
exercise\citep[see][and references therein]{urbsion}. Using the
models of hot stationary discs of \citet{wh98} to describe the
spectra of cold (and non-stationary) quiescent discs cannot be
expected to give relevant results. In fact since the observational
data were the UV spectra obtained by the IUE, it was not very
surprising that the best fits were obtained with no disc emission at
all: quiescent discs are not supposed to be UV emitters so that the
only source of UV radiation in such systems should be the white
dwarf.

The description of the quiescent state of the dwarf-nova cycle is
the Achilles heel of the widely accepted disc instability model
(DIM) according to dwarf-nova outbursts are due to a thermal-viscous
instability \citep[see][for a review]{l01}. For example while the
DIM predicts the quiescent disc to be optically thick (especially in
its outer regions), the observations of the quiescent dwarf-nova IP
Peg seem to suggest the opposite \citep{little01,ribeiro07}. Correct
modeling of the emission from quiescent discs is therefore an
obvious prerequisite to solving this thorny problem .

In general a full description of dwarf-nova outburst cycle should
include the radiation transfer equations. In practice this much too
costly and therefore unfeasible. However, instead of including one
can combine the DIM with the solutions of the radiative transfer
equations for the same disc parameters. The \citet[][hereafter
HMDLH]{hmdl98} version of the DIM in which the time-dependent radial
evolution equations use as input a pre-calculated grid of
hydrostatic vertical structures (a 1+1D scheme) is well suited to
such enterprize. The vertical structures are calculated using the
standard equations of stellar structure or the grey-atmosphere
approximation in the optically thin case. The angular-momentum
viscosity transport mechanism is described by the $\alpha$-ansatz of
\citet{ss73}. For a given $\alpha$, $M/R^3$ and effective
temperature $\Teff$ there exist a unique solution describing the
disc vertical structure. Such solutions which are very handy for
solving the time-dependent equations of the disc evolution cannot be
used to produce emission spectra. These can be, however, calculated
by using the same input parameters ($\alpha$, $M/R^3$ and $\Teff$)
in a radiative-transfer code on the condition that the vertical
solutions calculated by the two methods are the same. In this way
one can reproduce spectra of the whole cycle of a dwarf-nova
outburst.

The outline of the present article is as follows. In section
\ref{sec:model} we discuss the model used with special stress put on
marking the differences with original SW code on which it has been
based. In section \ref{sec:results} we present solutions obtained
for various types of physical set-ups. In subsection \ref{subsec:sc}
we compare our solutions for vertical structure with those obtained
with HMDLH code. Then in \ref{subsec:wh} we compare our disc spectra
with the UV spectra obtained by \citet{wh98}. In the following
subsection {subsec:non-stat} we present spectra of cold
non-equilibrium discs which in the DIM represent the quiescent state
of the dwarf-nova outburst cycle. Section \ref{sec:concl} contains
the discussion, and conclusion.

\section{Assumptions, equations and methods of solution}
\label{sec:model}

The structure of the code is based on the model of \citet[][]{sw91}.
The principal improvements consist in using (modern) line opacities
and in a different treatment of the convective energy transport. It
seems that our code is the only working accretion-disc
radiative-transfer code which includes convection self-consistently
and is able to handle dominant convective fluxes.

\subsection{Assumptions}

The  accretion disc is described as composed of concentric rings orbiting the
central gravitating body of mass $M$ with the Keplerian angular frequency
\be
\Omega_{\rm K}=\sqrt{\frac{GM}{R^3}}.
\label{eq:kepler}
\ee

We assume that the disc is geometrically thin, i.e.
\begin{equation}
\label{eq:geo_thin}
z_0 \ll R
\end{equation}
where $z_0$ is the disc height.
This assumption allows to decompose the disc equations into their vertical
and radial components. In this paper we will solve only the vertical
accretion-disc equations and obtain the corresponding emission spectra.
However, since the radiation of a steady-state accretion disc can be considered
as a sum of radiation from individual rings our results can directly applied
to such configurations. The same is true of quasi-stationary phases of
dwarf-nova outbursts. Also radiation from non-steady quiescent discs of
dwarf-nova stars can be treated as being the sum of the emission of individual
rings. This is true as long as radial gradients are much smaller than the
vertical ones. During the outburst of dwarf nova when the temperature and densities
fronts are propagating through the disc this assumption might no longer be valid.

\subsection{Equations}

The vertical structure of a ring at a distance $R$ from the center
is found by solving:

\begin{itemize}

\item The hydrostatic equilibrium equation
\begin{equation}
\frac{dP}{dz}=-\rho\,g_z=- \rho\,\Omega_K\,z,
\end{equation}

\item The energy balance equation
\be
\frac{dF_{\rm z}}{dz} = Q_{\rm vis}
=  \left(\frac{dF_{\rm z}}{dz}\right)_{\rm rad} + \left(\frac{dF_{\rm z}}{dz }\right)_{\rm conv},
\ee
where $F_{\rm z}$ is the energy flux in the vertical ($z$) direction.
We use the $\alpha$ prescription of \citet{ss73} to describe
the viscous energy generation
\begin{equation}
Q_{\rm vis}(z,R)=\frac{3}{2} \alpha \Omega_{\rm K} \frac{P}{\rho}.
\end{equation}

\item The radiative energy flux is given by
\begin{equation}
\label{eq:temp}
\begin{array}[t]{r}
\int_0^\infty\,\left(J(z,R,\lambda)-B(T(z,R),\lambda)\right)\kappa(\lambda) d\lambda \\
= -Q_{\rm vis}(z,R)
\end{array}
\end{equation}
where $J$ is the mean intensity, $B$ the Planck function and $\kappa(\lambda)$ is the
monochromatic mass absorption coefficient.

\item The radiative transfer equation is
\begin{equation}
\mathbf{n} \cdot \mathbf{{\nabla}} I=\kappa(S-I),
\end{equation}
where $I$ is the intensity, $S$ the source function and $\mathbf{n}$ is a unit vector in the ray direction.
\item Following HMDLH the convective energy flux is calculated by the method of \citet{p69}.

Whenever the radiative gradient
\be
\nabla_{\rm rad}\equiv \left(\frac{d\ln T}{d \ln P}\right)_{\rm rad}
\ee
is superadiabatic, the temperature gradient of the structure $\nabla$ is convective
($\nabla=\nabla_{\rm conv}$). The convective gradient is calculated in the
mixing length approximation, with the mixing length taken as $H_{\rm ml} =
\alpha_{\rm ml} H_P$, where $H_P$ is the pressure scale height:
\begin{equation}
H_P = {P \over \rho g_{\rm z} +(P \rho)^{1/2} \Omega_{\rm K}},
\end{equation}
The convective gradient is found from:
\begin{equation}
\nabla_{\rm conv} = \nabla_{\rm ad} + (\nabla_{\rm rad} - \nabla_{\rm ad}) Y
(Y+A)
\end{equation}
where $\nabla_{\rm ad}$ is the adiabatic gradient, and $Y$ is the solution of
the cubic equation:
\begin{equation}
{9 \over 4} {\tau_{\rm ml}^2 \over 3 +\tau_{\rm ml}^2} Y^3 + VY^2 + V^2 Y -V
= 0
\end{equation}
where $\tau_{\rm ml} = \kappa \rho H_{\rm ml}$ is the optical depth of
the convective eddies. It is assumed that the eddies are optically thick so that the convection
is  efficient, yet the contribution to the adiabatic gradient $\nabla_{\rm ad} + (\nabla_{\rm rad} - \nabla_{\rm ad}) Y
(Y+A)$ cannot be neglected. In those cases in which the convection reached the photosphere the effect on the weak lines must be noticeable. The coefficient $V$ is given by:
\begin{eqnarray}
V^{-2} & = & \left( {3 + \tau_{\rm ml}^2 \over 3 \tau_{\rm ml}}\right)^2 {g_{\rm
z}^2 H_{\rm ml}^2 \rho^2 C_P^2 \over 512 \sigma^2 T^6 H_P} \nonumber \\
& & \hspace*{3em} \times \left( {\partial
\ln \rho \over \partial \ln T} \right)_P (\nabla_{\rm rad} - \nabla_{\rm ad})
\end{eqnarray}
In our model we take $\alpha_{\rm ml} = 1.5$

\item The outer boundary condition is
\be
Q_{\rm vis}=\sigma T_{\rm eff}^4
\ee

In the case of a stationary disc
\begin{equation}
\dot{M} = 2 \pi R \int_{0}^{Z_0}\rho v_r dz =constant,
\end{equation}
where $v_r$ is the radial flow velocity
one has
\begin{equation}
Q_{\rm vis}=\frac{3}{8\pi\sigma}\left(\frac{GM_{\rm wd}\dot{M}}{R^3}\right)
\left(1-\left(\frac{R_{wd}}{R}\right)^{1/2}\right)
\end{equation}
but our solutions apply to any distribution of accretion rate $\dot M(r)$ in
a geometrically thin accretion disc.
\end{itemize}

\subsection{Input (opacities and  EOS)}

We have improved several features of the original SW code but the
use of modern opacities and especially of the line opacities is the
major improvement that we present here in some detail.

There are several ways of calculating opacities. They can be
obtained using either codes such as PHOENIX \citep{phoenix}, ``Atlas
12" \citep{kurucz} or from the database of the OP project
\citep{op}. The updated version of the SW code can use the opacities
from ``Atlas 12" but since calculating the opacities is time
consuming we decided to use tabulated data from the OP database. To
tabulate the data we used the data from the Opacity Project -
$OPCD\_3.3$ that were downloaded from the website at the Centre des
Donn\'{e}s de Strasbourg (CDS). For a given abundance mixture the
subroutines $mx.f$, $mixv.f$, $opfit.f$ and $mixz.f$ allowed us
(with some modifications) getting the tabulated opacities  for the
wavelengths chosen as our basic grid. For the emerging spectra from
every ring we use 10,000 wavelengths starting from $100${\AA} up
to $10^6${\AA}.  The grid interval varies between $\Delta \lambda
=0.25${\AA} for the most important zone such as the range between
$850-2300${\AA} and $1-2${\AA} used for wavelengths shorter than
$800${\AA} or above $2300${\AA}. For wavelengths longer than
$5000${\AA}, $\Delta \lambda$ varies logarithmicaly  starting
at $2${\AA} and getting up to $200${\AA} for the very long
wavelengths. All the data were tabulated using a grid for the
temperature $T$ and electron density $Ne$. The indices $ite$ and
$jne$ are defined by
\begin{equation}
\begin{array}[t]{l}
    ite=40x\log(T),  \ \ jne=4x\log(Ne) \\
    \triangle(ite)=\triangle(jne)=2
    \end{array}
\end{equation}

The mesh in the tables is calculated for $140<ite<320$ therefore the
lowest temperature that can be used in the code is $3160 \rm K$. Let
us stress that the opacities in the present work do NOT take into
account the global line broadening due effects such as velocities,
\citep[expansion opacity, see ][]{sw05}, the inclination of the
disc or microscopic such as microturbulence. Consequently, the lines
provide only an indication of the presence of a given ion and cannot be
used, so far, for abundance determinations etc. These broadenings
will be included in the code in the near future. Therefore the
results presented here should be treated as more qualitative than
quantitative in the sense that for the moment we will not discuss the
ratios between the emission or absorption lines but rather their
presence or absence in the spectrum.

\begin{figure}
 \includegraphics[width=\columnwidth]{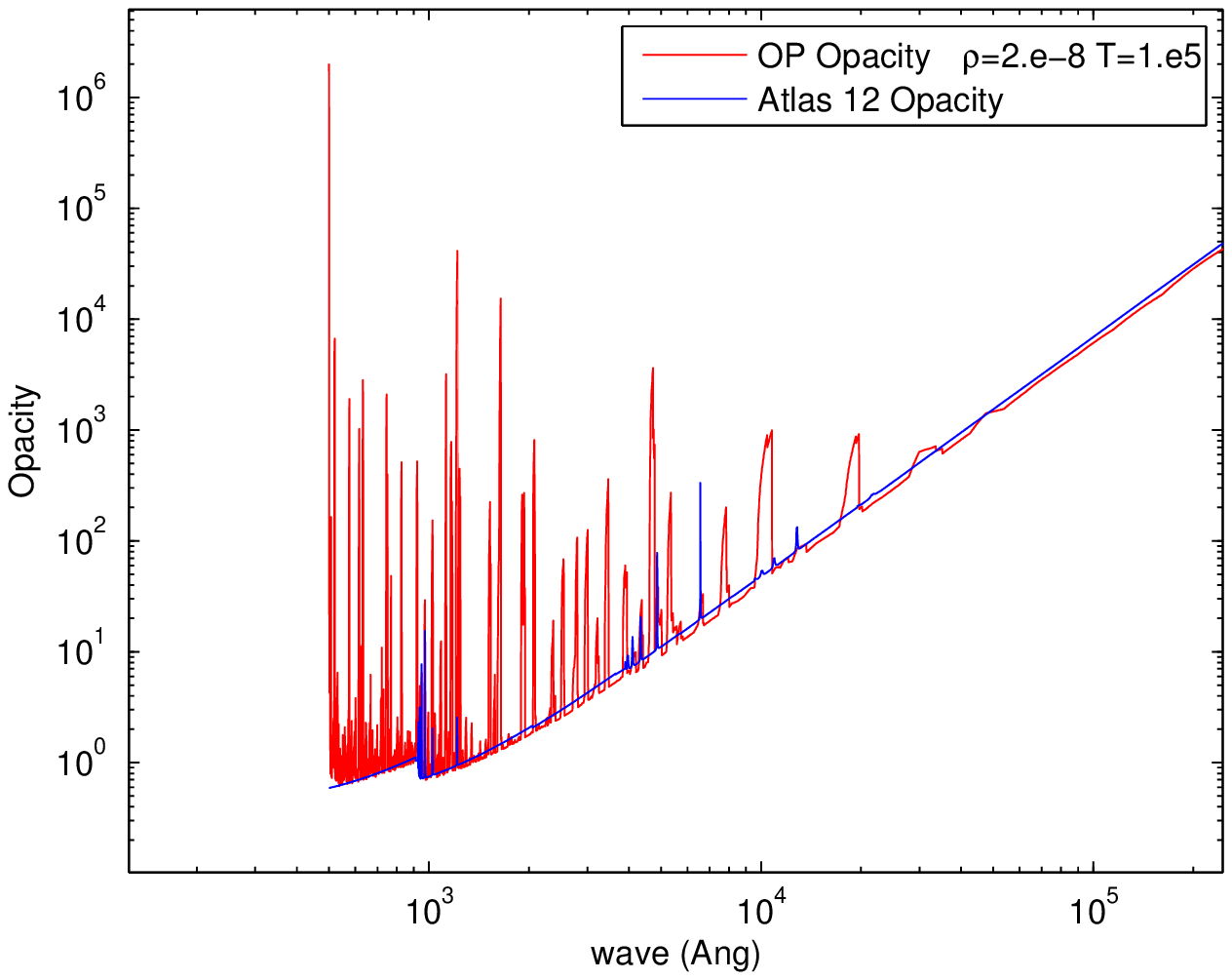} \\
 \includegraphics[width=\columnwidth]{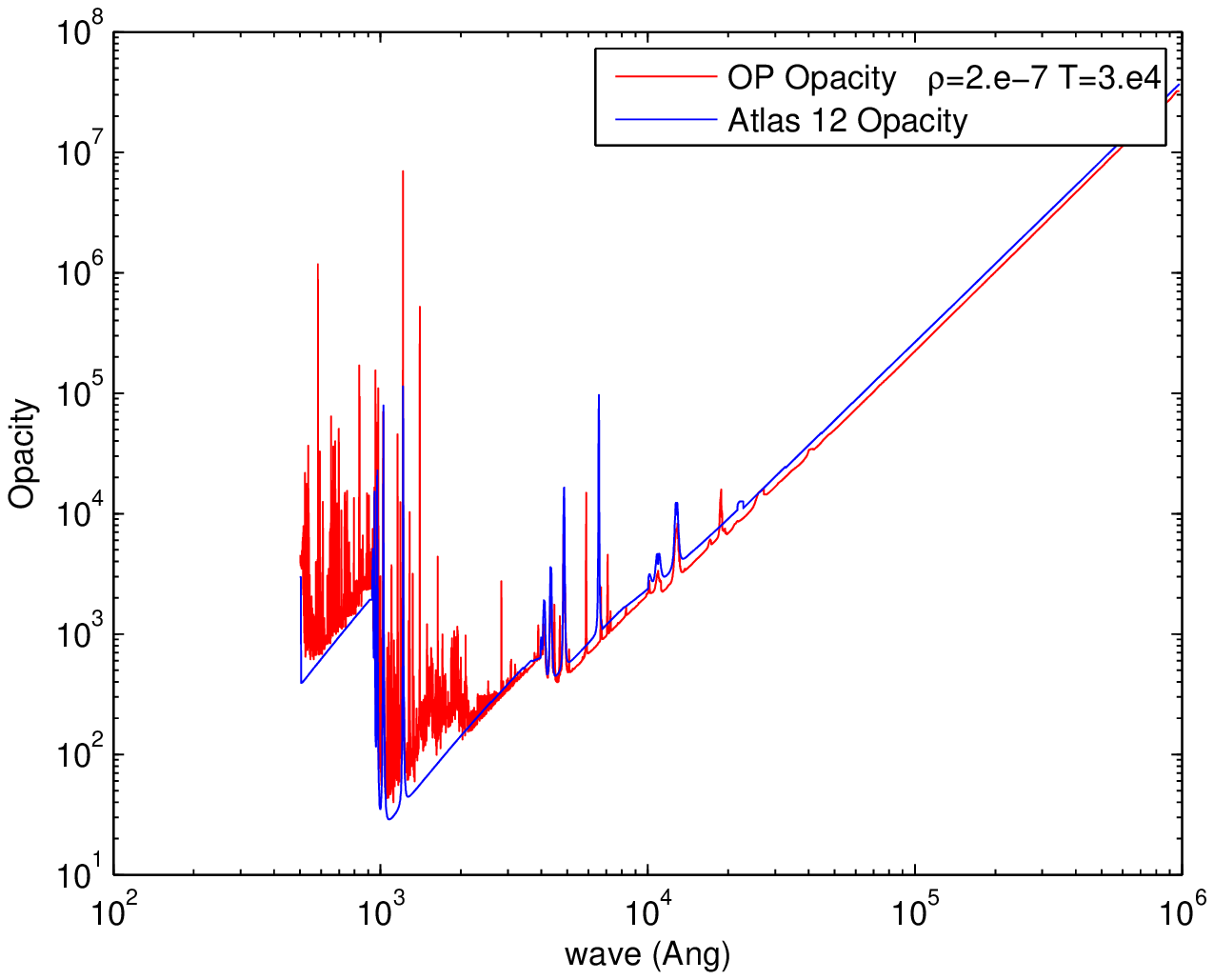}
 \caption{Comparision between the opacities as function of wavelength obtained
  from Atlas 12 and those obtain from the OP for T= and $\rho=$ (upper figure)
  and T= and $\rho=$ (lower figure). The Atlas 12 opacities are the continuum opacities
  with  only H and He lines, while the OP opacities are for solar mixture }
 \label{fig:opacity}
\end{figure}

The reason for allowing the code to work either with OP tabulated
data or with ``Atlas 12" was to test the tabulated opacities. In
Figure \ref{fig:opacity} we show the comparison of opacities as
function of wavelength calculated for $T=10^5$K and
$\rho=2\times10^{-8}$ g cm$^{-3}$, and $T=3\times10^4$K and
$\rho=2\times 10^{-7}$g cm$^{-3}$ using either ``Atlas 12" or the
OP. The ``Atlas 12" continuum opacities used were for a solar
mixture with the addition of only  H and He lines. The opacities
from the OP Project are for also for mixture of solar abundance
(which includes the lines and the continuum).

The Rosseland mean opacities are taken from OPAL tables for solar
composition. While solving the radiative transfer equation we
checked for the consistency between the Rosseland mean opacity
obtaining from the solution of the radiative transfer equation and
the value from the OPAL table. The equation of state  is
interpolated from the tables of \cite{FGV}. The  equation of state
used in the OP project is different than the one use by Kurucz or in
OPAL. We are aware of the this problem and its effect on the
opacities but we will not discussed it here.

\subsection{The method of solution}

The basic idea behind the SW code was to couple the hydrostatic with
the radiative-transfer equations. Therefore the code iterates the
two equations simultaneously.  The iteration continues until a
structure is found to which both the solution for the  hydrostatic
and the radiative-transfer equations converge.

The temperature profile is determined from Eq. (\ref{eq:temp}). The
temperature $T(z,R)$ in this equation is solved by expressing $J$ in
terms of $B$ using the radiative transfer equation (for detail
description see \cite{kw84},\cite{w81}). As mentioned above we
assumed that each radial ring is independent of its neighboring rings
hence there is only a vertical radiative flux. The radiative
transfer equation is solved in the two stream approximation. $I^+$
represent the specific intensity in the outward direction and
$I^-$ is the specific intensity in the inward direction. The
radiative equation is then written as:
\begin{eqnarray}
\pm\frac{dI^{\pm}(R,z,\lambda)}{dz} &=& -(\kappa(\lambda)+\sigma(\lambda))I^\pm(R,z,\lambda)
\\
& + & \sigma(\lambda)J(R,z,\lambda)+\kappa(\lambda)B(T(R,z),\lambda)
\nonumber
\end{eqnarray}
where $\sigma(\lambda)$ is the scattering coefficient. We use the
same N grid  points as in the hydrostatic calculation. The boundary
conditions are: no incident flux ($I^{-}_{j,1}=0$) and at the
equatorial plane the  upward and downward specific intensities are
equal ($I^{+}_{j,N}=I^{-}_{j,N}$), where j designates the ring
number.

When solving the vertical-structure equations of an accretion disc
the main problem is that contrary to stellar atmospheres  the exact
height of the photosphere is unknown a-priori. Because of gravity
increasing with height a change in the position of the photosphere
is followed by a change of the gravitational acceleration which
leads to a change in the entire vertical structure  of the disc. One
has to remember that the upper part of the disc is optically thin
and may include, depending on the assumptions, a non negligible
energy production. However, since this optically thin region is a
poor radiator and absorber a small energy production there might
have large effect on the final temperature structure. Once the line
opacities are taken into account  the importance of tuning the disc
height is crucial since otherwise the outgoing radiative flux might
not be equal to the total energy flux of the ring. Consequently the
great sensitivity of the emission lines to the details of the
temperature at the edge of the photosphere might be used to
investigate the properties of the energy-release (``viscosity")
mechanism.

In view of the above we approach the problem in the following way:
we guess first an initial temperature--optical depth relation and assume
an initial height $z_0$.  The optical depth in the disc uppermost
point is chosen to be $10^{-6}$, and this assumption is checked
again once the final structure is obtained. With the help of the
T($\tau$) relation we can integrate the hydrostatic equation from
$z_0$ down to $z=0$. One should emphasize that $\tau$ in the current
work is the Rosseland mean and not the optical depth at 5000{\AA} as
in SW. When the height $z_0$ is the correct height, the total energy
generation rate in the ring calculated from the energy equation is
equal to the total luminosity of that particular ring. If this is
not the case we alter $z_0$ keeping the T($\tau$) relation fixed,
until the this condition is satisfied. Once this iteration is solved
the hydrostatic model is consistent with the energy generation but
not yet with the radiation field.

We then turn to solving the radiative transfer equation using
the disc height, the pressure and the density structure obtained
from the hydrostatic iteration. This structure is kept fixed
during the iteration  for the radiative field.  We now solve the
radiative transfer equation and obtain a new $T(\tau)$ relation,
where new values of the temperature are assigned to old values of the
optical depth. If the new $T(\tau)$ relation from the radiative
iteration agrees (to a chosen degree of accuracy) with the old
$T(\tau)$ relation from the hydrostatic iteration the
hydrostatic structure is consistent with the radiation field. If
not we iterate again with the new $T(\tau)$ relation.

\begin{figure}
 \includegraphics[width=\columnwidth]{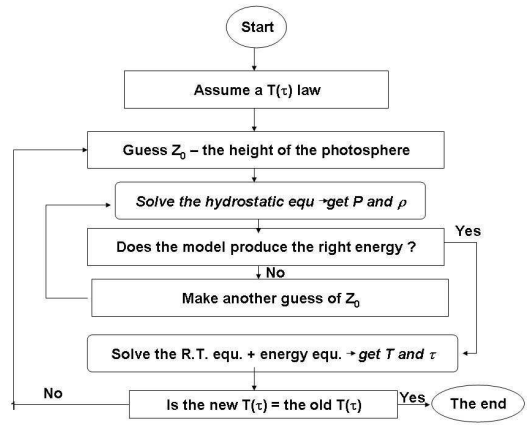}\\
 \caption{The iteration scheme}
 \label{fig:iteration}
\end{figure}

To summarize we have 3 basic iteration loops (see figure \ref{fig:iteration}) in the code :
\begin{enumerate}
  \item Calculating the hydrostatic structure using a given $T(\tau)$ relation.
  The output results are the disc height, pressure and density structure.
  \item Solving the radiative transfer for a given disc height, P and $\rho$.
  The output result is a new $T(\tau)$ relation.
  \item Repeating the previous two iterations until the $T(\tau)$ relation
  from the hydrostatic part is equal to the $T(\tau)$ relation from the radiative
  transfer equation.
\end{enumerate}

Only after these iterations converge  the emerging flux from the
disc is calculated using the radiative transfer equation.

As mentioned before the entire disc is divided into a series of
concentric rings and their width is determined according to their
distance from the WD. All rings which are  further away than $2\Rwd$
are assumed to have a  width of $1\Rwd$, below $2\Rwd$ the width of
the ring is taken to be $0.05\Rwd$, which allows a detailed
calculation of the emission of the rings near the boundary layer.

The vertical structure of the ring itself is divided into 100 $z$
points  (100 layers) and we use 10000 wavelengths. The above
procedure is repeated for each ring. The emerging fluxes from the
rings are then integrated  in order to get the total flux from the
entire disc according to:
\begin{equation}
\label{totflux}
    F_{\rm total}=\int_{\Rin}^{\Rout} F(R)RdR
\end{equation}

\section{Results}
\label{sec:results}

\subsection{Comparison with HMDLH vertical structures}
\label{subsec:onering}

As we mention in the introduction  a 1+1D scheme  was developed by
HMDLH to describe the time-dependent   evolution of the dwarf nova
outbursts. The scheme uses as input a pre-calculated grid of
hydrostatic vertical structures. For a given $\alpha$, $M/R^3$ and
effective temperature $\Teff$ there exist a unique solution
describing the vertical structure of a disc in thermal equilibrium.
HMDLH code cannot be used to produce emission spectra. These can be,
however, calculated by using the same input parameters ($\alpha$,
$M/R^3$ and $\Teff$) in the present code (hereafter ILHS) on the
condition that the vertical solution calculated by the two methods
are the same. In this way one can reproduce spectra of almost the
whole cycle of a dwarf-nova outburst. This procedure makes sense
only if the structures calculated by the two methods are
practically the same.

\begin{figure}
 \includegraphics[width=\columnwidth]{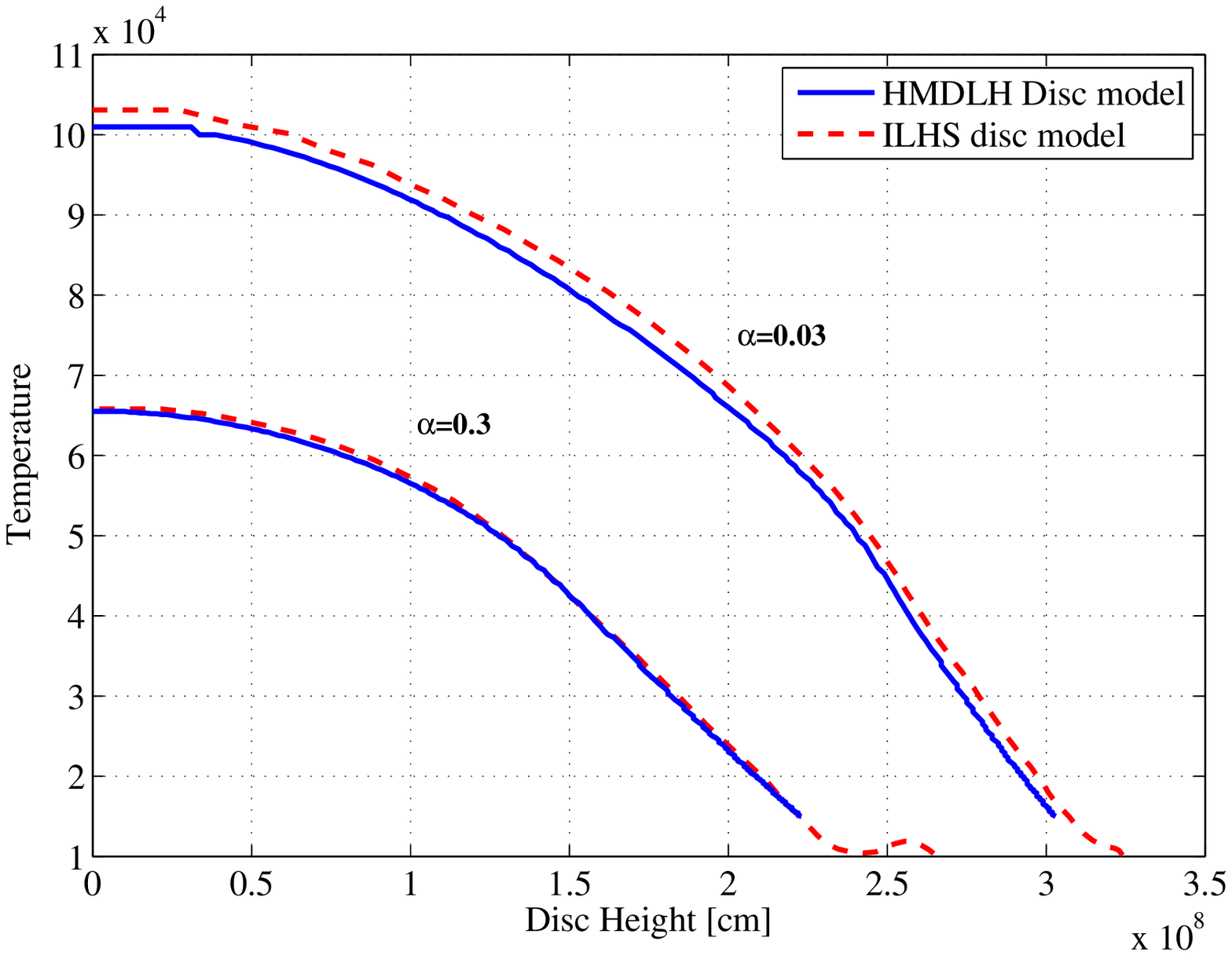} \\
 \includegraphics[width=\columnwidth]{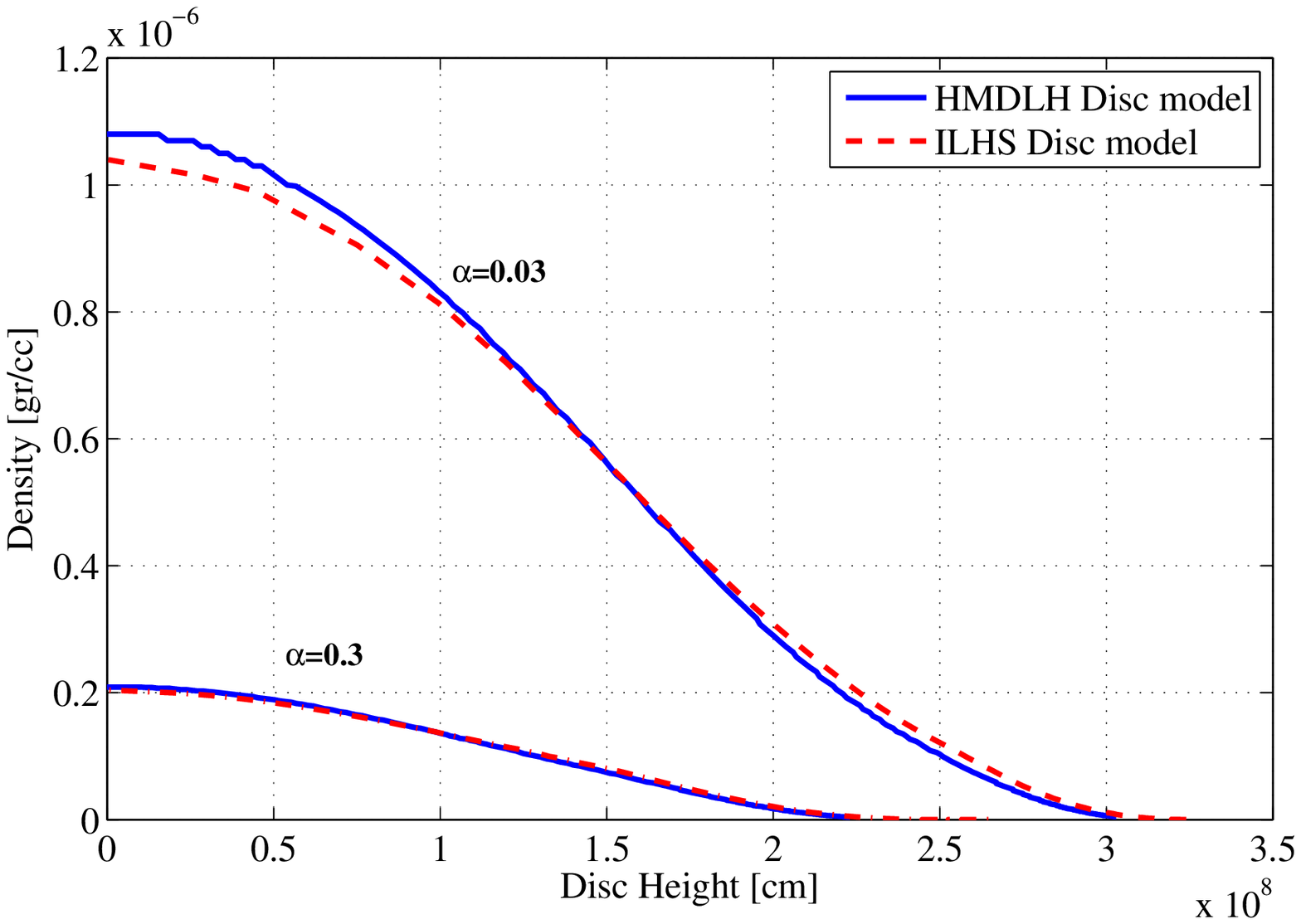}
 \caption{Comparison between the temperature and density profile at
 R=5$\Rwd$ and $\Teff$=$15000$K calculated by the HMDLH disc model
 and the present disc model taking into account the radiative transfer equation
 for $\alpha=0.3$ and $\alpha=0.03$.}
 \label{fig:teff_15000}
\end{figure}

\begin{figure}
 \includegraphics[width=\columnwidth]{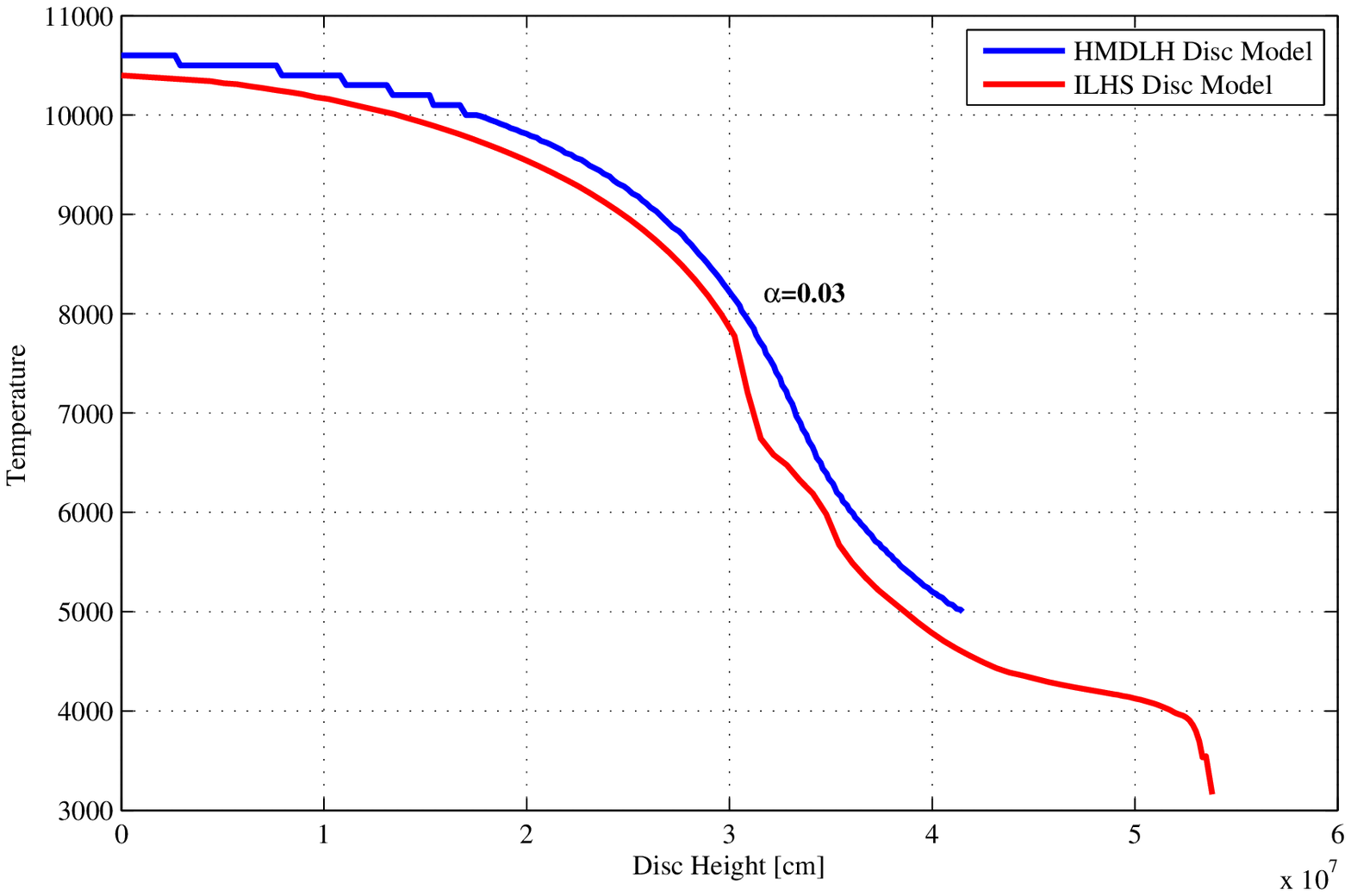} \\
 \includegraphics[width=\columnwidth]{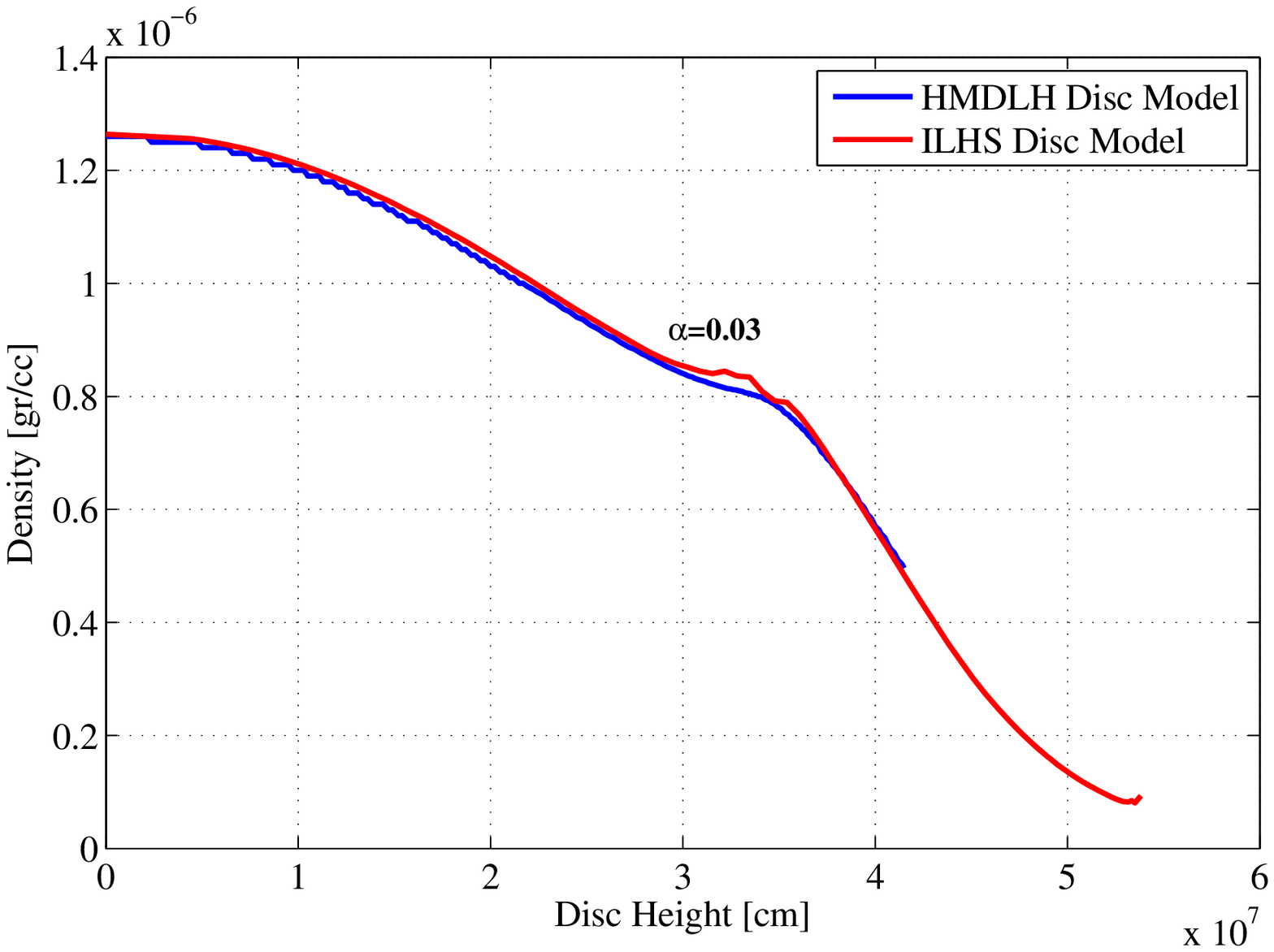}
 \caption{Comparison between the temperature and density profile at
 R=5$\Rwd$ and $\Teff$=$5000$K, $\alpha=0.03$ calculated by the HMDLH disc model
 and the present disc model taking into account the radiative transfer equation.}
 \label{fig:teff_5000}
\end{figure}

In order to compare the solutions obtained by HMDLH with those
calculated with  ILHS we show in Figures \ref{fig:teff_15000} and
\ref{fig:teff_5000}  the vertical structures for hot and cold discs
with effective temperatures $15000$K and $5000$K respectively. The
mass of the white dwarf is $0.6M_{\odot}$, the ring radius is
$5\Rwd$ where the white dwarf radius was calculated according to
\cite{nau72} mass-radius relation.We used two values of the
viscosity parameter $\alpha=0.03$ and $0.3$. One sees a perfect
agreement for the $\alpha=0.3$, while the minor differences for  the
$\alpha=0.03$ are of the order of $3\%$.

\subsection{S-curves}
\label{subsec:sc}

\begin{figure}
 \includegraphics[width=\columnwidth]{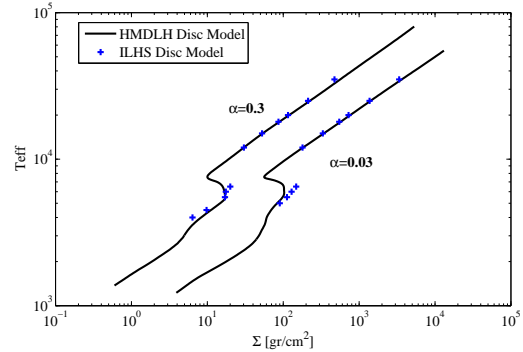}
 \caption{\ $\Sigma-\Teff$ curves for $R=\Rwd$, $\Mwd=0.6\Msun$,
 and $\alpha=0.03$ and 0.3. The solid line is the HMDLH model
 while + represent the results from the ILHS disc calculation.}
 \label{fig:scurve}
\end{figure}

For a given mass and distance from the center the thermal equilibria
of accretion discs can be represented as a $T_{\rm
eff}(\Sigma)$-relation. These equilibria forms an S shape on the
$(\Sigma,T_{\rm eff})$ plane. Figure \ref{fig:scurve} shows an
example`s of $\Sigma-\Teff$ S curves obtained with both the HMDLH
code and the radiative-transfer calculation of the ILHS. $\alpha$ is
constant and the radius is $5R_{\rm wd}$. Each point on the S-curve
represents a thermal-equilibrium at that radius. The upper branch of
the S-curve is the hot stable branch, while the lower branch where
the temperature is below $\simeq 6000$K, represent the cold stable
solutions. The middle branch of the S-curve corresponds to thermally
unstable equilibria.

The agreement between the two codes is very good especially on the
upper branch of the S-Curve. However, the disc at the lower turning
knee of the S-curve (in the unstable zone) is fully convective. Here
we were unable to calculate the structure with the full radiative
transfer equation code. The degree of accuracy that was chosen for
the $T-\tau$ to converge in these calculations was $1\%$. But since
near the lower turning knee of the S-curve the disc is almost
completely convective and any small change in the initial conditions
such as the initial optical depth will affect the structure of the
disc, we assume that the acceptable error are of the order of few
percent. The agreement between the two codes on the lower stable
branches of the S-curve is very good.

\subsection{The single ring spectrum}
\label{subsec:1ring}

\begin{figure}
 \includegraphics[width=\columnwidth]{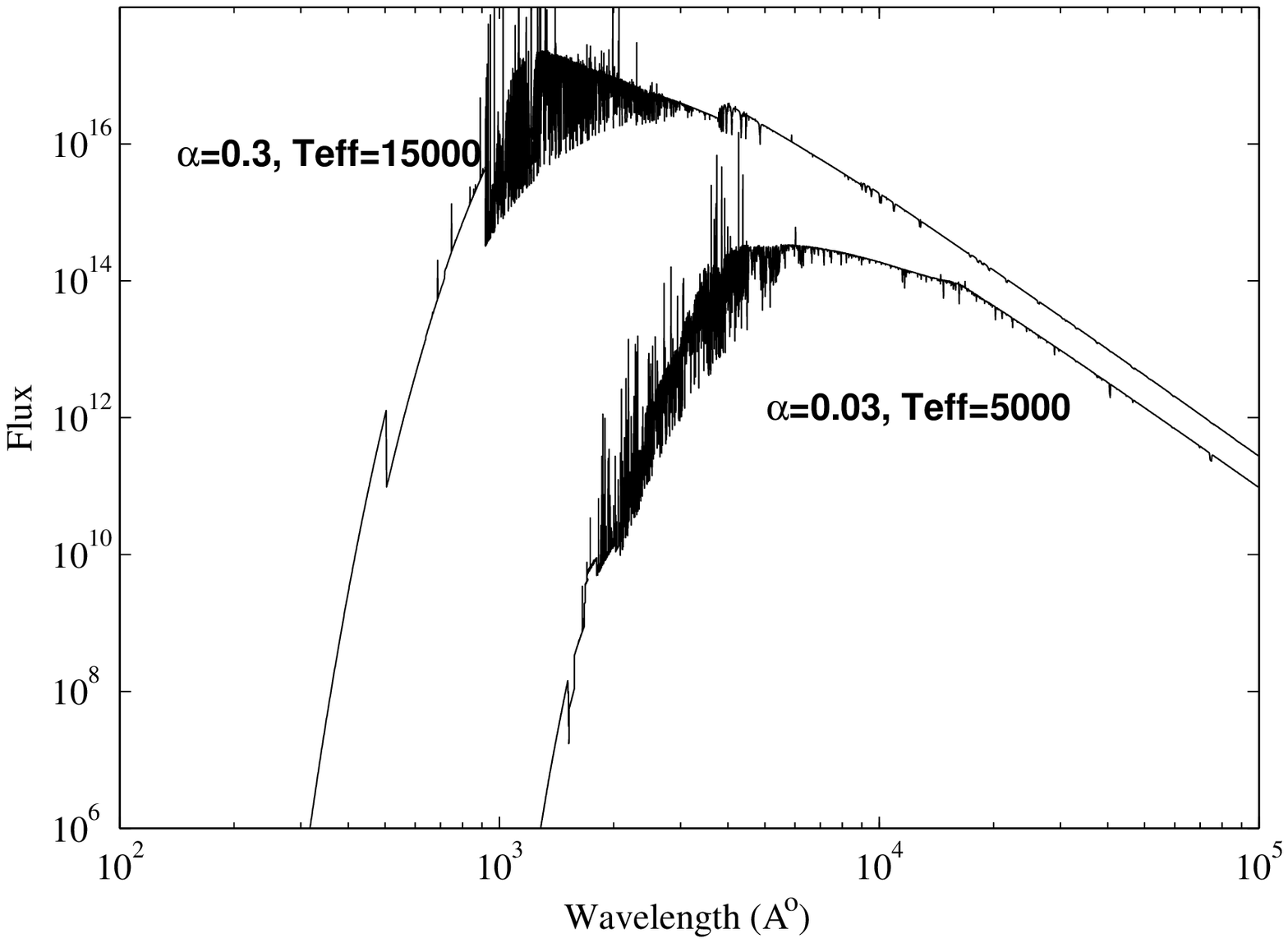}\\
 \caption{The spectra  obtained from a single-ring calculation for
 cold ($\alpha=0.3$,$\Teff=15000$K) and hot disc ($\alpha=0.03, \Teff=5000$K).
 The ring parameters are $\Rwd=5\Rsun$ and $\Mwd=0.6\Msun$.}
 \label{fig:flux_comp}
\end{figure}

\begin{figure}
 \includegraphics[width=\columnwidth]{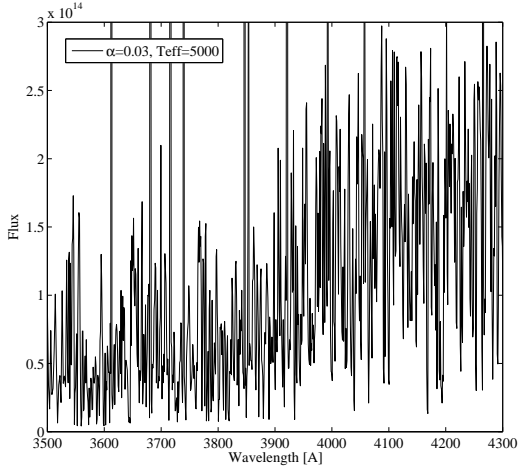}\\
 \caption{The flux  at the wavelength between
 $3500-4000${\AA}, for cold disc ($\alpha=0.03, \Teff=5000$K).
 The spectrum is dominated by emission lines.}
 \label{fig:a3543_003}
\end{figure}

\begin{figure}
 \includegraphics[width=\columnwidth]{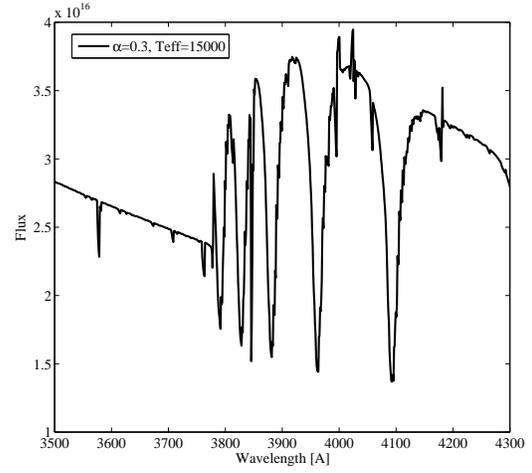}\\
 \includegraphics[width=\columnwidth]{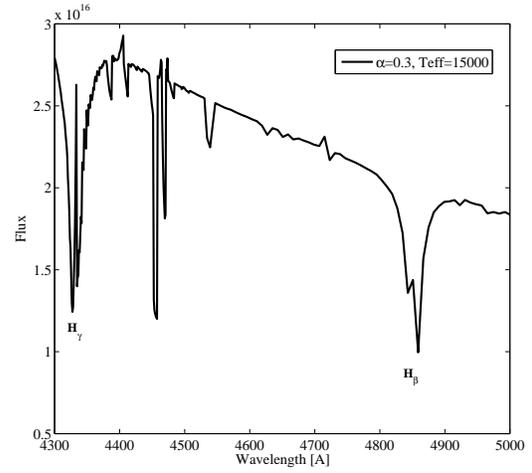}\\
 \includegraphics[width=\columnwidth]{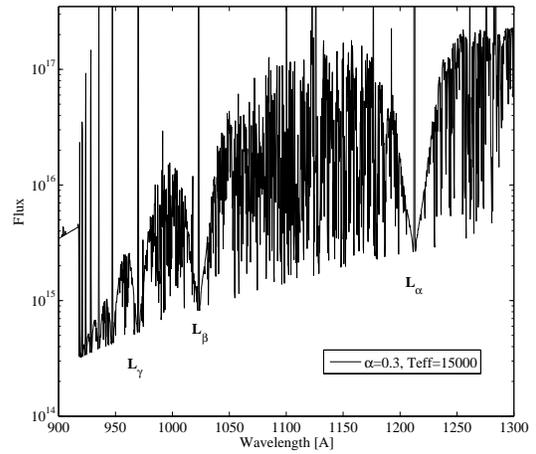}\\
 \caption{The flux as function of wavelength,
 for hot disc ($\alpha=0.3,\Teff=15000$K). The spectrum is
 dominated by absorption lines and in some cases narrow emission
 line in the central of the absorption ``well" }
 \label{fig:flux_comp1}
\end{figure}

In Figures \ref{fig:flux_comp}, \ref{fig:a3543_003} and
\ref{fig:flux_comp1} we present the spectra obtained from
single-ring calculations at $\Rwd=5\Rsun$ for $\Mwd=0.6\Msun$. The
spectra (in arbitrary units) were calculated for a cold disc with
$\alpha=0.03$ $\Teff=5000$K and a hot disc with $\alpha=0.3$,
$\Teff=15000$K.  The spectrum of the cold disc is dominated by
narrow emission lines while the spectrum of the hot disc is
dominated by wide absorption line. In the hot disc spectrum
(\ref{fig:flux_comp1}) we can identify the Balmer lines (the
$4100{\AA},4340{\AA},4860 {\AA}$ lines). The absorption lines have
various structures : in some cases such as the Lyman lines or the
$H_\gamma$ line in hot disc, there is an emission line in the center
of the absorption line. We should emphasize again that one must be
careful when interpreting the results at this point, since in this
paper the opacities do not include line broadening and are not
taking into account the Gaussian instrumental broadening function
that affects and smears most of the fine lines. Such broadening can
change or even eliminate some of the fine structure of the lines
that is seen in the figures.

\subsection{Stationary disc solutions for high accretion rates and $\alpha=0.3$ }
\begin{figure}
 \includegraphics[width=\columnwidth]{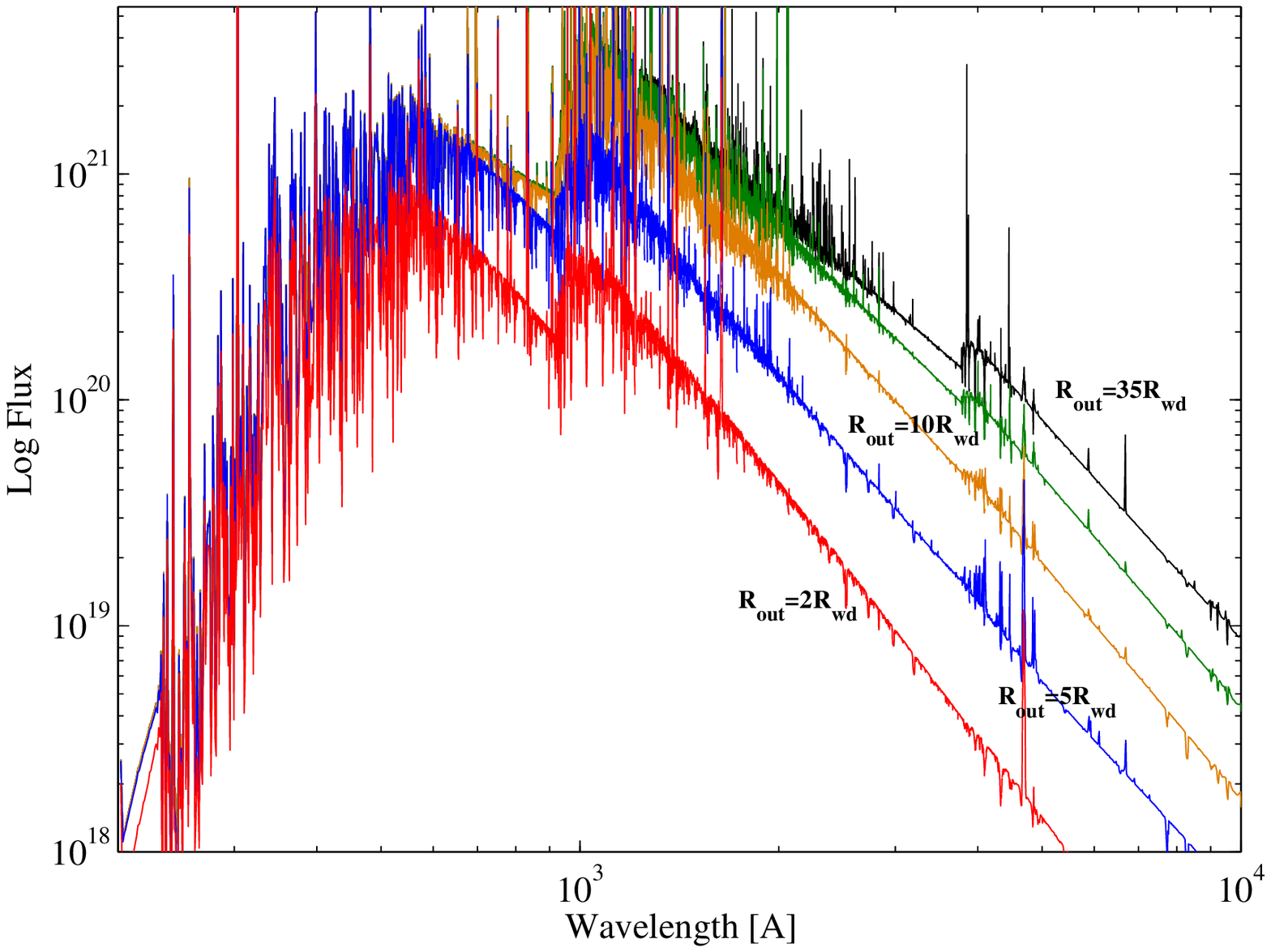} \\
 \caption{The integrated disc spectrum as function of the disc radii for a
 hot disc $\alpha=0.3, \Mwd=0.8\Msun$ }
 \label{fig:total_flux_03_17}
\end{figure}

In  Figures \ref{fig:total_flux_03_17} we present an example for the
total spectrum of a hot disc  $\alpha=0.3, \Mwd=0.8 \Msun$ and
accretion rate of $10^{17}\rm g\,s^{-1}$. We show spectra for
increasing disc area - from the inner part of the disc (the inner
$2\Rwd$ area) up to the largest outer radius calculated - $35\Rwd$.
One can clearly see the different area in the disc in which
different type of emission is formed. For example: the Balmer jump
which will dominate the total flux from the disc is formed on the
outer part of the disc, while the inner part of the disc is
dominated only by few absorption lines.

\subsection{Comparison with Wade \& Hubeny hot accretion disc spectra}
\label{subsec:wh}

\citet[][hereafter WH]{wh98} calculated a large grid of  far- and
mid-ultraviolet spectra  (850-2000 {\AA}) of the integrated light
from steady-state accretion disks in bright cataclysmic variables.
All their models are for disc atmospheres which  are optically thick
and are viewed from a distance of $100 \rm pc$.

\begin{figure}
 \includegraphics[width=\columnwidth]{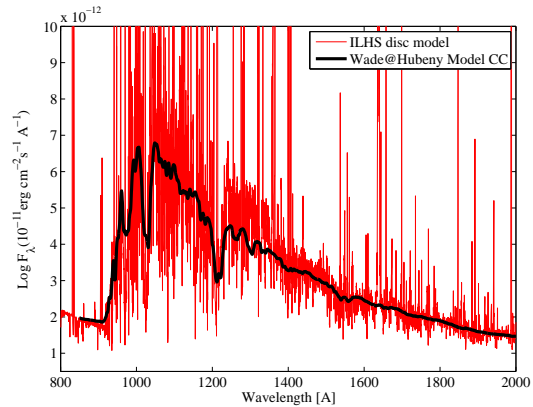}
 \caption{Comparison between Wade and Hubeny model cc spectrum (thick black line)
 and the spectrum from our disc model (grey line) for $\alpha=0.3$
 and $\log\dot{M}=-8.5 \Msun \rm yr^{-1}$. }
 \label{fig:wadedc}
\end{figure}

Figure \ref{fig:wadedc} shows a comparison between a WH accretion
disc spectrum (model cc) and a spectrum obtained with our model. The
model parameters were are: $\Mwd=0.8 \Msun$ and $\log\dot{M}=-8.5
\Msun \rm y^{-1}$. In our disc model we used  and the same number of
rings as in their paper. The viscosity prescription in WH is
different  from the $\alpha$ ansatz used in our model. Their
viscosity is based on Reynolds number approach \citep{kh86} . A value of $Re=5000$
was assumed for all radii. 
We estimated that $\alpha=0.3$ corresponds to the parameters
of WH model cc.

WH spectra are smoother than our spectra since their models have
been convolved with a Gaussian instrumental profile but one can
expect that in our spectrum most of the narrow emission lines will
disappear after such convolution. In general both spectra are
similar.

\subsection{Quiescent dwarf-nova disc: on-stationary solution $\alpha=0.03$, $\Teff=5000$.}
\label{subsec:non-stat}

Dwarf-nova quiescent discs are non-stationary, i.e. $\Mdot \neq
constant$ since they are filling-up getting ready for the next
outburst. The whole disc must be cold, its temperature everywhere
must be lower than the critical instability temperature. In general,
for most of the quiescence the temperature profile is roughly flat.
We modeled as an example the spectrum of such a disc assuming
$\Teff= 5000$K (which is a bit high for a real disc since the
critical temperature is close to that value).

\begin{figure}
\includegraphics[width=\columnwidth]{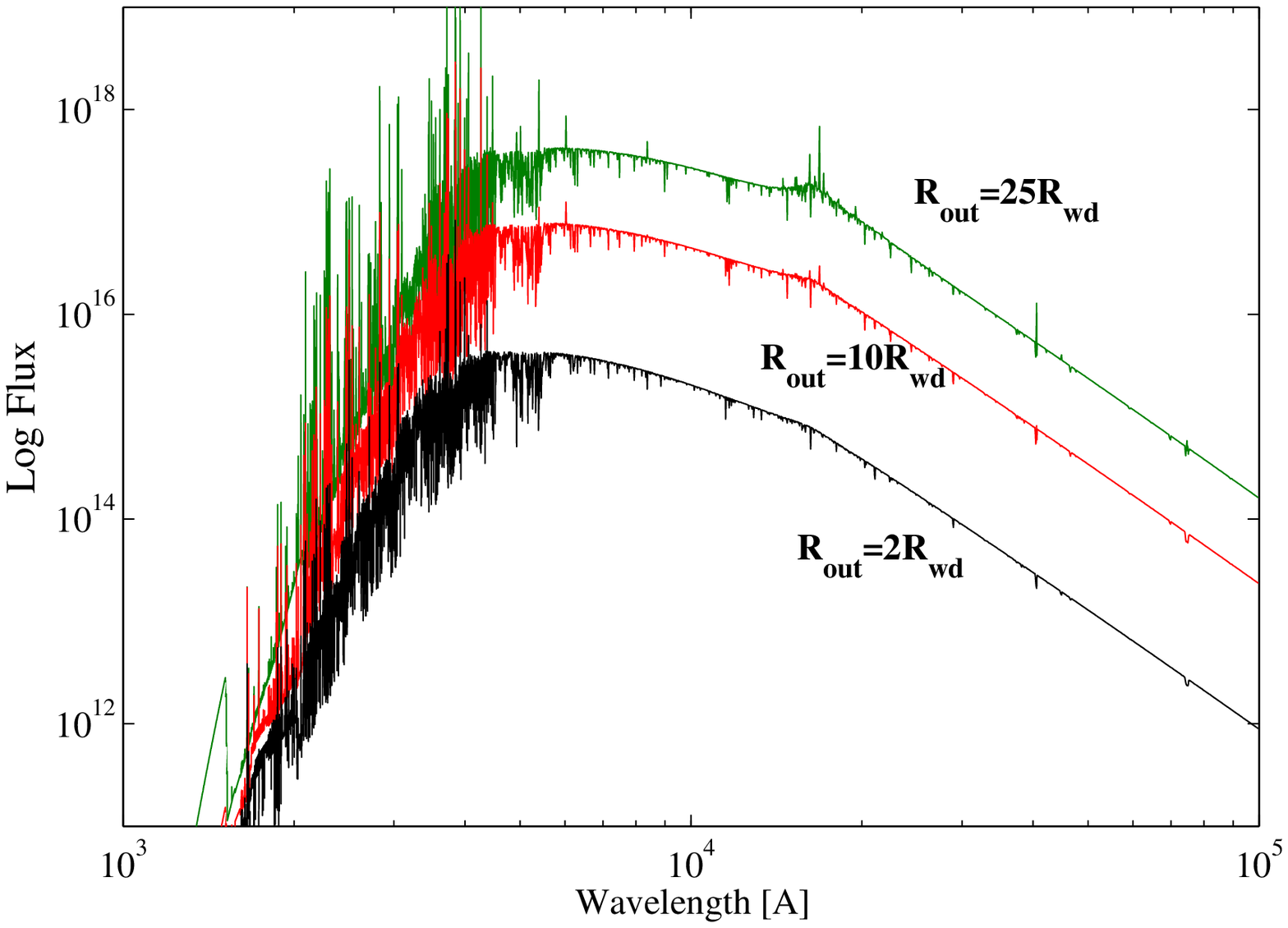} \\
 \includegraphics[width=\columnwidth]{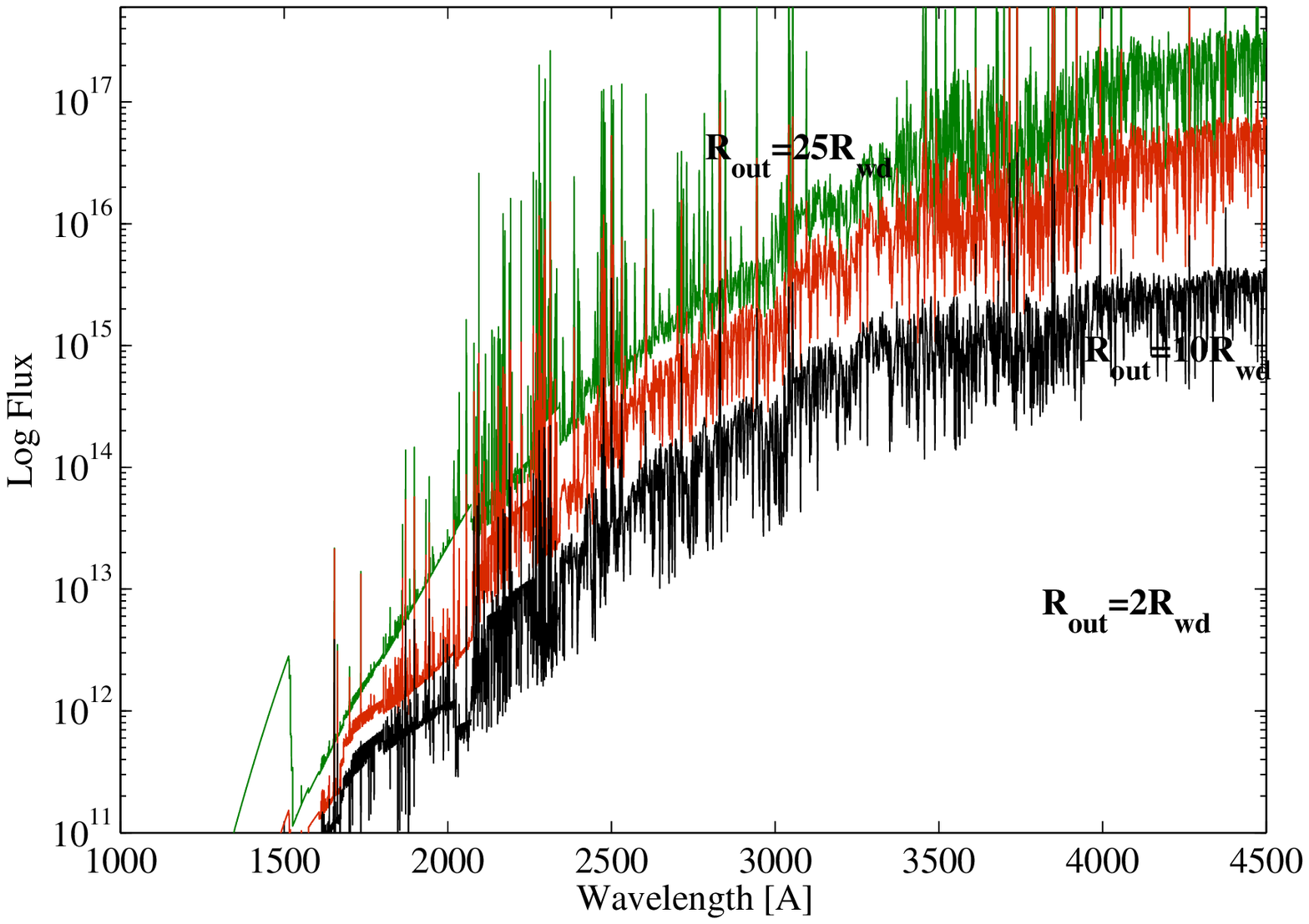}
 \caption{The integrated disc spectrum as function of the disc radii for a
 cold disc $\alpha=0.03,\Teff=5000 \Mwd=0.6\Msun$}
 \label{fig:total_flux_003_5000}
\end{figure}

The total flux for an outer radius of $25\Rwd$ is shown in figure
\ref{fig:total_flux_003_5000}. The spectrum between
$3000-10000${\AA} is nearly a flat as observed, no Balmer jump is
present. As can be seen from figure \ref{fig:total_flux_003_5000}
for this particular calculation, there is no contribution from the
disc spectrum in the UV range.

\section{Future work}
\label{sec:concl}

In the future we will produce a grid of disc spectra will be used to
make comparisons between observations and the model spectra. In
particular will focus on the quiescence disc and the effect of the
irradiation by the white dwarf and the formation of a corona.
Combining ILHS with the time dependent HMDLH code will provide a
tool to study the dwarf nova outburst cycle.

The variations in the shapes  of  selected  lines that are formed in
different parts of the disc will be studied with the purpose to
develop  spectral-disc-tomography.

\section*{ Acknowledgments:} We thank C. Zeippen and F. Delahaye for their valuable help with the opacities. We also would like to thank R. Wehrse and I. Hubeny for suggestions and  discussions.



\end{document}